\documentclass[12pt,fleqn]{article}
\usepackage{amsmath, amssymb,amsthm,cite}
\usepackage{fullpage}

\title{Invertible Mappings of Nonlinear PDEs to Linear PDEs Through
Admitted Conservation Laws}
\author{ 
Stephen Anco \\
\small \emph{Department of Mathematics, Brock University, St. Catharines, ON Canada L2S 3A1}\\
George Bluman \\
\small \emph{Department of Mathematics, University of British Columbia, Vancouver, V6T 1Z2 Canada}\\
Thomas Wolf \\
\small \emph{Department of Mathematics, Brock University, St. Catharines, ON Canada L2S 3A1}\\}

\newtheorem{theorem}{Theorem}

\newtheorem{prop}{Propostion}
{\theoremstyle{definition}

} {\theoremstyle{definition}
\newtheorem{definition}{Definition}}

\def \zeroall {\setcounter{theorem}{0} \setcounter{equation}{0} \setcounter{remark}{0} \setcounter{definition}{0} \setcounter{prop}{0} }

\def\beq{\begin{equation}}
\def\eeq{\end{equation}}
\def\barr{\begin{array}{ll}}
\def\earr{\end{array}}

\newcounter{tabnum}

\newcounter{def_} \newcounter{rem_}  

\begin{document}

\maketitle

\begin{abstract}

An algorithmic method using conservation law multipliers is introduced that yields necessary and
sufficient conditions to find invertible mappings of a given nonlinear
PDE to some linear PDE and to construct such a mapping when it
exists. Previous methods yielded such conditions from admitted point
or contact symmetries of the nonlinear PDE.  Through examples, these
two linearization approaches are contrasted.
\end{abstract}



\section{Introduction}
\setcounter{equation}{0} 

In previous work \cite{b01,b02,b03}, algorithmic methods
were presented to determine whether or not there exists an invertible
mapping of a given nonlinear partial differential equation (PDE) to
some linear PDE through computing the point or contact symmetries 
admitted by the nonlinear PDE. 
Moreover, it was also shown how to construct such an
invertible mapping, when one exists, from the admitted symmetries.

In this paper, we present an alternative algorithmic method for
mapping a nonlinear system of PDEs invertibly to some linear system of
PDEs by using admitted conservation laws of the nonlinear system. 
Any linear system possesses an infinite set of conservation law
multipliers satisfying its adjoint system. This feature of linear
systems will be exploited to detect whether or not a given nonlinear
system can be linearized by an invertible transformation and, when
such a linearization mapping exists, to obtain the explicit form of
the linearizing transformation from an associated conservation law
identity.

In particular, using the determining equations for 
conservation law multipliers, 
one can find whether or not an invertible
linearization mapping will exist, and also find the adjoint of a target
linear system as well as the transformation of the independent variables 
in the mapping whenever an invertible linearization is possible \cite{b04}. 
Most importantly, the new
work presented in this paper will show that a conservation law
identity coming from the multiplier equations of an augmented system
consisting of the given nonlinear system and the adjoint of the linear
target system can be used to determine the 
transformation of the dependent variables 
(in addition to that of the independent variables) 
in the invertible linearization mapping explicitly. 
This work thus allows one to detect and construct 
linearization mappings completely by the use of
algorithmic methods for obtaining multipliers of conservation laws 
\cite{b05,b06,b07,b08,a}.

Our results generalize earlier computational work \cite{w}
in which a fully automatic algorithm was given using 
the computer algebra program {\sc Crack/Conlaw} \cite{b14}
to detect and construct linearizations 
in the case when the mapping only transforms the 
dependent variables of the given nonlinear PDE system. 
The work in \cite{w} also gave an extension of this algorithm to find 
partial linearizations that consist of factoring out 
a linear differential operator from all PDEs in a given system
(with the operator being of lower order than the highest order of derivatives
in the system). 

To begin we summarize two well-known theorems 
characterizing invertible mappings. 
Our notation follows that in \cite{b11},
with the conventions that Latin indices $i,j,k$ will run $1$ to $n$
(labeling independent variables), 
and Greek indices $\nu,\mu,\lambda$ and $\alpha,\beta,\gamma,\sigma,\tau$
will run $1$ to $M$ and $1$ to $m$ 
(labeling PDEs and dependent variables), respectively. 
Throughout we use upper/lower case notation to distinguish 
arbitrary functions $U,V$ from functions $u,v$ given by solutions of a PDE system.

Let ${\bf R}\{x,u\}$ denote a given $k$th-order nonlinear system of
$M$ PDEs with $n$ independent variables $x = (x_1 ,\ldots ,x_n )$ and
$m$ dependent variables $u = (u^1,\ldots ,u^m)$. In particular, let
the PDEs in ${\bf R}\{x,u\}$ be given by
\begin{equation}
\label{eq1} 
G^\nu [u] = G^\nu (x,u,\partial_x u,\ldots ,\partial_x^ku) = 0,\quad \nu = 1,\ldots ,M,
\end{equation}
where $\partial_x^ju$ denotes $j$th-order partial derivatives of $u$ with respect to $x$.

Let ${\bf S}\{z,w\}$ denote a $k$th-order linear target system of $M$\ PDEs
with $n$ independent variables $z = (z_1 ,\ldots ,z_n )$ and $m$
dependent variables $w = (w^1,\ldots ,w^m)$,
\begin{equation}
\label{eq2} 
H^\nu [w] = H^\nu (z,w,\partial_z w,\ldots ,\partial_z^k w) = 0,\quad \nu = 1,\ldots ,M,
\end{equation}
where $\partial_z^j w$ denotes $j$th-order partial derivatives of $w$ with respect to $z$.

Note that for common PDE systems such as 
heat or diffusion equations and wave equations, 
as well as more general parabolic, hyperbolic and Hamiltonian equations, 
and elliptic equations, 
we will have $m=M$. 
However, for complete generality we allow $m\neq M$ hereafter, 
as would occur for instance if one considers 
a PDE system of dynamical evolution equations with differential constraints. 

Now if one seeks an invertible mapping $\mu$ of 
a given nonlinear PDE system (\ref{eq1}) to a linear target PDE system (\ref{eq2}), 
then the following two theorems 
show that the mapping is restricted to being a contact transformation 
when (\ref{eq1}) and (\ref{eq2}) both contain a single dependent variable
($m = 1$; which we will call the {\it scalar case})
or to a point transformation 
when (\ref{eq1}) and (\ref{eq2}) each contain two or more dependent variables
($m \ge 2$; the {\it multicomponent case}). 

\begin{theorem}[B\"{a}cklund \cite{b09}]
If $u=u^1$ is a scalar $(m = 1)$, then a mapping $\mu $ 
defines an invertible transformation from
$(x, u, \partial_x u, \ldots , \partial_x^p u)$ -- space to
$(z,w,\partial_z w,\ldots ,\partial_z^pw)$ -- space for \emph{any}
fixed $p$ if and only if $\mu$ is a \emph{one-to-one contact transformation} 
of the form 
$z = \phi (x,u,\partial_xu)$, $w = \psi (x,u,\partial_xu)$, 
$\partial_z w = \rho(x,u,\partial_xu)$, 
which is subject to the contact condition $D_x \psi = \rho D_x \phi$ where $D_x$ denotes total derivatives with respect to $x$.
\end{theorem}

\begin{theorem}[M\"{u}ller and Matschat \cite{b10}]
If $u=(u^1,\ldots,u^m)$ has two or more components $(m \ge 2)$, then a mapping $\mu$ defines an invertible transformation from
$(x,u,\partial_x u,\ldots ,\partial_x^p u)$ -- space to
$(z,w,\partial_z w,\ldots ,\partial_z^p w)$ -- space for \emph{any}
fixed $p$ if and only if $\mu$ is a \emph{one-to-one point
transformation} of the form
$z = \phi (x,u)$, $w = \psi (x,u)$.
\end{theorem}

For formulating necessary and sufficient conditions 
for the existence of such mappings between PDE systems, 
we will consider only the natural situation where 
the given nonlinear system (\ref{eq1}) and the linear target system (\ref{eq2})
contain the same number $M$ of PDEs with the same differential order $k$,
and no further assumptions will be required 
(such as local solvability, or involutivity). 
But for the computational implementation of our results 
it is important that the given system of PDEs (\ref{eq1})
be consistent in the sense that all integrability conditions are
satisfied modulo these PDEs. 
(To make this precise, 
one would have to prescribe an elimination algorithm 
which requires the definition of a total ordering of derivatives and of monomials built from them, as discussed in \cite{w}.)

The rest of this paper is organized as follows. 
In section two, we review the method presented in \cite{b01,b02,b03} 
to linearize nonlinear PDEs through admitted symmetries. 
In section three, we present the new method to
linearize nonlinear PDEs through admitted conservation law multipliers
and outline the computational steps involved in this method. 
As examples to illustrate and contrast both methods, 
in section four we consider linearizations of 
Burgers' equation, a pipeline flow equation, and a nonlinear telegraph system. 
We make some closing remarks in section five.

\section{Use of Symmetries to Construct Linearizations}\zeroall

Necessary and sufficient symmetry conditions are stated in \cite{b01,b02,b03}
to determine whether or not a given nonlinear system ${\bf R}\{x,u\}$ of PDEs
(\ref{eq1}) can be transformed to some linear target system ${\bf
S}\{z,w\}$ of PDEs (\ref{eq2}) by an invertible mapping $\mu $.  
These conditions provide a symmetry-based algorithmic procedure to determine
if they hold for a given nonlinear system of PDEs and
(when these conditions do hold) 
a second symmetry-based algorithmic procedure to
construct such an invertible mapping $\mu $. 
To apply these methods to a given nonlinear system of PDEs, it is
unnecessary to know a specific linear target system of PDEs. In
particular, the linear target system (when one exists) will arise from
computing the admitted point or contact symmetries 
(Lie group of point or contact transformations) 
of the given nonlinear system (first procedure). 
Moreover, these admitted symmetries yield equations to
construct a specific mapping $\mu $ (second procedure).

The starting point is that a linear system ${\bf S}\{z,w\}$ of PDEs, given by
\begin{equation}
\label{eq5} L[z]w = g(z)
\end{equation}
as defined here in terms of a linear operator L[$z$], with any
inhomogeneous term $g(z)$, is completely characterized by its admitted
infinite-parameter set of point symmetries
\begin{equation}
\label{eq6} {\bf X} = f(z) \frac{\partial }{\partial w},
\end{equation}
where $f(z)$ is any function satisfying the related linear
homogeneous system
\begin{equation}
\label{eq7} L[z]f = 0.
\end{equation}

A point transformation is mapped to another point transformation under
any specific point transformation $(x,u) \rightarrow (z,w)$,
and a contact transformation is mapped to another contact transformation
(which could be a point transformation) under any specific contact
transformation $(x,u,\partial_xu) \rightarrow (z,w,\partial_zw)$
(which could be a point transformation). Hence an invertible point
transformation mapping a nonlinear system ${\bf R}\{x,u\}$ of PDEs to
some linear system ${\bf S}\{z,w\}$ must map each point symmetry
admitted by ${\bf R}\{x,u\}$ to a point symmetry admitted by ${\bf
S}\{z,w\};$ an invertible contact transformation mapping a nonlinear
scalar PDE system ${\bf R}\{x,u\}$ to some linear scalar PDE system
${\bf S}\{z,w\},$ must map each contact symmetry admitted by ${\bf
R}\{x,u\}$ to a contact symmetry (which could be a point symmetry)
admitted by ${\bf S}\{z,w\}$. 
Moreover, this mapping must be an
isomorphism, i.e., there is a one-to-one correspondence between the
point (contact) symmetries admitted by ${\bf R}\{x,u\}$ and those
admitted by ${\bf S}\{z,w\}$. 
Consequently, if there exists an
invertible transformation that maps the nonlinear PDE system ${\bf R}\{x,u\}$
to some linear PDE system ${\bf S}\{z,w\},$ then ${\bf R}\{x,u\}$ 
must necessarily admit an infinite set of point (contact) symmetries.

The following two theorems on mappings of nonlinear PDEs to linear PDEs
were presented and proved in \cite{b01,b02,b03}. 

\begin{theorem}[Necessary conditions for the existence of 
an invertible linearization mapping]
If there exists an invertible mapping $\mu$ of a given nonlinear
system ${\bf R}\{x,u\}$ of PDEs \eqref{eq1} 
to some linear system ${\bf S}\{z,w\}$ of PDEs \eqref{eq2}, then
in the multicomponent case (i.e. $m \ge 2$):
\begin{enumerate}
\item $\mu$ is a point transformation of the form
\beq \label{eq8} 
z_i = \phi_i (x,u),\quad i = 1,\ldots ,n, \qquad
w^\sigma = \psi ^\sigma (x,u),\quad \sigma = 1,\ldots ,m;
\eeq
\item ${\bf R}\{x,u\}$ admits an infinite-parameter Lie group of point symmetries given by an infinitesimal generator
\begin{equation}
\label{eq9}
{\bf X} = \xi_i (x,u)\frac{\partial }{\partial x_i } + \eta ^\tau(x,u)\frac{\partial }{\partial u^\tau}
\end{equation}
{with infinitesimals of the form} 
\beq \label{eq10} 
\xi_i =
\alpha _{i\sigma }(x,u) F^\sigma (x,u), \quad 
\eta^\tau = \beta_\sigma^\tau(x,u) F^\sigma (x,u), 
\eeq 
where 
$\alpha_{i\sigma}$, $\beta_\sigma^\tau$ 
($i = 1,\ldots ,n;\sigma,\tau = 1,\ldots ,m$) 
are specific functions of $x$ and $u$, 
and where $F = (F^1,\ldots ,F^m)$ 
is an arbitrary solution of some linear system of PDEs
\begin{equation}
\label{eq11} 
L[X]F = 0
\end{equation}
in terms of some linear differential operator $L[X]$ and specific
independent variables 
$X = (X_1 (x,u),\ldots ,X_n (x,u))=(\phi_1,\dots,\phi_n)$.
\end{enumerate}
\end{theorem}

\begin{theorem}[Sufficient conditions for the existence of an invertible 
linearization mapping] 
Suppose a given nonlinear system ${\bf R}\{x,u\}$ of PDEs (\ref{eq1}) 
admits infinitesimal point symmetries \eqref{eq9} of the form \eqref{eq10}
involving an arbitrary solution $F(X)$ of a linear system
\eqref{eq11} with specific independent variables 
$X = (X_1(x,u),\ldots ,X_n (x,u))$. 
If the $m$ first order scalar linear homogeneous PDEs
\begin{equation}
\label{eq12} 
\alpha _{i\sigma}(x,u)\frac{\partial \phi }{\partial x_i } 
+ \beta_\sigma^\tau(x,u) \frac{\partial \phi}{\partial u^\tau} = 0,
\quad \sigma = 1,\ldots ,m,
\end{equation}
whose coefficients are formed from \eqref{eq10} 
have $\phi_1=X_1 (x,u),\ldots ,\phi_n=X_n (x,u)$ as $n$ functionally independent solutions, 
and if the $m^2$ first order linear inhomogeneous PDEs
\begin{equation}
\label{eq13} 
\alpha _{i\sigma}(x,u)\frac{\partial \psi^\tau}{\partial x_i} 
+ \beta_\sigma^\gamma(x,u) \frac{\partial \psi^\tau}{\partial u^\gamma} 
= \delta_\sigma^\tau ,
\quad \tau,\sigma = 1,\ldots ,m,
\end{equation}
(where $\delta _\sigma^\tau$ is the Kronecker symbol) 
have a particular solution
$\psi = (\psi ^1(x,u),\ldots ,\psi ^m(x,u)),$ then the mapping $\mu$ defined by
\beq \label{eq14} 
z_i = X_i (x,u),\quad i = 1,\ldots ,n,\qquad
w^\sigma = \psi^\sigma(x,u),\quad \sigma = 1,\ldots,m ,
\eeq
is invertible and transforms ${\bf R}\{x,u\}$ to the linear system ${\bf S}\{z,w\}$ of PDEs given by
$L[z]w = g(z),$
for some inhomogeneous term $g(z)$.
\end{theorem}

It is easy to see that Theorems 2.1 and 2.2 also hold 
in the scalar case (i.e. when $m=1$) 
if one seeks to transform nonlinear scalar PDEs 
to linear scalar PDEs only by a point transformation. 
Furthermore, these theorems in the scalar case 
can be extended to the more general possibility of
seeking invertible mappings of 
nonlinear scalar PDEs to linear scalar PDEs by contact transformations. 
For details, see \cite{b01,b02}.

\section{Use of Conservation Law Multipliers to Construct Linearizations}\zeroall

Now we present a new method for finding mappings that transform
a nonlinear system of PDEs invertibly to some linear system of PDEs.
Our method uses admitted conservation law multipliers of the nonlinear system
and is based on two theorems that are the
conservation law multiplier-based counterparts of the symmetry-based Theorems~2.1 and~2.2. 

The first new theorem gives necessary conditions that conservation law multipliers must satisfy in order for 
a given nonlinear system ${\bf R}\{x,u\}$ of PDEs (\ref{eq1}) 
to have an invertible mapping to some linear target system ${\bf S}\{z,w\}$ 
of PDEs (\ref{eq2}) by a point transformation. 
As in the case of Theorem~2.1, 
the necessary conditions yield an algorithmic procedure to
find a specific linear target system as well as its independent variables.

The second new theorem gives sufficient conditions for the existence of the mapping. In particular, as in the case of Theorem 2.2, 
these conditions yield an algorithmic procedure to find 
the components of an explicit point transformation giving 
an invertible mapping to the linear target system.

The extension of these two theorems to include contact transformations 
will be discussed in section 3.2.

\subsection{Necessary and Sufficient Linearization Conditions}
Recall, a conservation law of a system ${\bf R}\{x,u\}$ of PDEs (\ref{eq1}) is a divergence expression
$D\Upsilon_i[u]/Dx_i$ that is required to vanish for all solutions $u(x)=(u^1(x),\ldots,u^m(x))$ of the given PDE system. 
Conservation laws are nontrivial if and only if the conserved fluxes $\Upsilon_i[u]$ do not have the form of curls 
$D\Theta_{ij}[u]/Dx_j$, with $\Theta_{ij}[u]=-\Theta_{ji}[u]$, for every solution $u(x)$.

\begin{definition}
\textit{Conservation law multipliers} for a system ${\bf R}\{x,u\}$ of PDEs (\ref{eq1}) are a set of factors
$\{\Lambda_\nu(x,U,\partial_x U,\ldots,\partial_x^\ell U)\}$ 
$(\nu=1,\ldots,M)$ such that, 
for arbitrary functions $U = (U^1(x),\ldots ,U^m(x))$, 
\begin{equation}
\label{eq26} 
\Lambda_\nu[U] G^\nu[U] \equiv D\Upsilon_i[U] / Dx_i
\end{equation}
holds as an identity for some functions $\Upsilon = (\Upsilon_1[U],\ldots,\Upsilon_n[U])$,
where each factor $\Lambda_\nu[U]$ is non-singular for all solutions $U=u(x)$
of the given PDE system (\ref{eq1}).
\end{definition}

It is well known that multipliers will yield all nontrivial conservation laws (to within the
addition of curls to the conserved fluxes) admitted by a 
given PDE system ${\bf R}\{x,u\}$ if it is of Cauchy-Kovalevskaya type.

For any PDE system ${\bf R}\{x,u\}$, its admitted conservation law multipliers satisfy the determining equations
\begin{equation}
\label{deteq} 
E_{U^\sigma}(\Lambda_\nu[U]G^\nu[U]) = 0, \quad \sigma= 1,\ldots,m, 
\end{equation}
given by the Euler operators 
\[
E_{U^\sigma} = \frac{\partial}{\partial
U^\sigma} - \frac{D}{Dx_i}\frac{\partial}{\partial U^\sigma_{x_i}} 
+\frac{D^2}{Dx_iDx_j}\frac{\partial}{\partial U^\sigma_{x_ix_j}} +\ldots 
+(-1)^k\frac{D^k}{Dx_i\ldots Dx_j}\frac{\partial}{\partial U^\sigma_{x_i\ldots x_j}} 
+\ldots 
\]
which annihilate divergence expressions.  
One can solve these determining equations by an algorithmic
procedure \cite{b08,w} 
analogous to that for solving the determining equations
for point or contact symmetries in evolutionary form.
(The similarity is made explicit in the work in \cite{b06,b07} 
which shows how to reformulate (\ref{deteq}) as a kind of adjoint of
the symmetry determining equations for Cauchy-Kovalevskaya PDE systems.)

The starting point, leading to the formulation of an invertible mapping between 
nonlinear and linear PDE systems through conservation law
multipliers, is that any linear operator $L$ and its adjoint
operator $L^*$ satisfy a conservation law identity.
Specifically, consider a $k$th-order linear PDE system $L[z]w = 0$,
denoted by ${\bf S}\{z,w\}$ with dependent variables $w=(w^1,\ldots,w^m)$ 
and independent variables $z=(z^1,\ldots,z^n)$. 
In particular, let the linear PDEs be given by
\begin{equation}
\label{eq27} 
L_\alpha ^\nu [z]w^\alpha = 0, \quad \nu = 1,\ldots ,M
\end{equation}
in terms of linear operators
\footnote{We freely raise and lower indices for convenience here in displaying 
the adjoint relation between $L$ and $L^*$.}
\begin{equation}
\label{eq28} 
L_\alpha ^\nu [z] =
b_{\nu \alpha}(z) + b_{\nu \alpha i}(z)\frac{\partial }{\partial z_i } 
+ \cdots
+ b_{\nu \alpha i_1 \cdots i_k}(z)
\frac{\partial^k}{\partial z_{i_1 } \cdots \partial z_{i_k } }.
\end{equation}
The corresponding adjoint linear system, $L^*[z]v = 0,$ is given by
\beq \label{eq29}
\barr
L_\nu^{*\alpha}[z]v^\nu 
= \displaystyle b_{\mu\alpha}(z)v^\mu 
-\frac{\partial}{\partial z_i}\Big(b_{\mu\alpha i}(z)v^\mu\Big) + \cdots  
+ ( - 1)^k\frac{\partial ^k}{\partial z_{i_1 } \cdots \partial z_{i_k } }
\Big(b_{\mu\alpha i_1 \cdots i_k}(z)v^\mu \Big)
\\\qquad\qquad
= 0,  \ \ \alpha = 1,\ldots,m, 
\earr\eeq
where $L^{*\alpha}_\nu[z]$ is the linear operator defined through formal 
integration by parts of the operator $L^\nu_\alpha[z]$ with respect to $z$. 
Then for arbitrary functions $V(z)=(V^1(z),\ldots,V^M(z))$ 
and $W(z) = (W^1(z),\ldots ,W^m(z)),$ which can be viewed as multipliers $\{V^\mu(z)\}$
and $\{W^\alpha (z)\}$ for the augmented linear system consisting of the linear system (\ref{eq27}) and the adjoint system
(\ref{eq29}), there is a conservation law identity
\begin{equation}
\label{eq31} 
D\Upsilon_i / Dz_i = 
\delta_{\nu\mu} V^\mu L_\alpha^\nu [z]W^\alpha - 
\delta_{\alpha\beta}W^\beta L_\nu^{*\alpha} [z]V^\nu 
\end{equation}
holding for some specific functions $\Upsilon_i[V,W]$ 
that have a bilinear dependence on the multipliers and their derivatives.
(An explicit expression for the fluxes $\Upsilon_i[V,W]$ is given in \cite{b}.)
Here both $\delta_{\nu\mu}$ and $\delta_{\alpha\beta}$ are Kronecker symbols.

\begin{prop}
Suppose a given nonlinear system ${\bf R}\{x,u\}$ of PDEs (\ref{eq1}) 
can be mapped into some linear system ${\bf S}\{z,w\}$ of PDEs (\ref{eq27})
by some point transformation (\ref{eq8}). 
Then the nonlinear and linear PDEs will be related through 
a set of factors $\{Q_\nu^\mu(x,U,\partial_x U)\}$ such that 
\begin{equation}
\label{eq32} 
Q_\nu^\mu [U]G^\nu [U] = L_\alpha^\mu [z]W^\alpha ,\quad \mu = 1,\ldots ,M,
\end{equation}
for arbitrary functions $U(x) = (U^1(x),\ldots ,U^m(x))$, 
where the functions $W(z) = (W^1(z),\ldots ,W^m(z))$ are
given by the linearization mapping 
\[ 
z = \phi (x,U(x)),\quad W(z) = \psi (x,U(x)) . 
\]
(Explicit expressions for $Q_\nu^\mu$ can be readily derived
from the mapping formulae given in \cite{f}.)
Moreover :
\begin{enumerate}
\item 
This mapping will define an invertible point transformation iff 
$\{\phi_i(x,U(x))\}$ ($i=1,\ldots,n$) 
are functionally independent
and $\{Q_\nu^\mu(x,U(x),\partial_x U(x))$ ($\nu,\mu=1,\ldots,M$) 
are non-degenerate,
namely, 
$\det(D\phi_i(x,U)/Dx_j)\neq 0$
and also $\det( Q_\nu^\mu(x,U,\partial_x U) )\neq 0$. 
\item 
Under such a point transformation
the conservation law identity (\ref{eq31}) becomes
\begin{equation}
\label{eq33} D\Gamma_i / Dx_i = (\delta_{\lambda\mu}V^\lambda Q_\nu^\mu [U]G^\nu [U] 
- \delta_{\alpha\beta}W^\beta L_\nu^{*\alpha} [z]V^\nu )J[U],
\end{equation}
where $D\Gamma_i / Dx_i$ is related to $D\Upsilon_i / Dz_i$
by the non-vanishing Jacobian factor 
\begin{equation}
\label{eq34} 
J(x,U,\partial_x U) 
= \left| {\frac{Dz}{Dx}} \right| 
= \det \left( {\frac{D\phi_i(x,U(x))}{Dx_j }} \right).
\end{equation}
(See \cite{b11} for the explicit expression for $\Gamma_i$ 
in terms of $\Upsilon_i$.) 
\end{enumerate}
\end{prop}

This proposition leads to the following two theorems for linearization from
admitted conservation law multipliers, 
which are the counterparts to Theorems~2.1 and 2.2 for linearization 
through admitted point symmetries.

\begin{theorem}[Necessary conditions for the existence of an invertible 
linearization mapping]
If there exists an invertible point transformation \eqref{eq8} 
for $m\geq 1$ under which 
a given $k$th-order nonlinear system ${\bf R}\{x,u\}$ of PDEs \eqref{eq1} 
is mapped to some linear system ${\bf S}\{z,w\}$ of PDEs (\ref{eq27}), 
then ${\bf R}\{x,u\}$ must admit conservation law multipliers of the form 
\begin{equation}
\label{eq35} 
\Lambda_\nu [U] = \delta_{\lambda\mu}v^\lambda(X) Q_\nu^\mu [U] J[U],
\end{equation}
where 
$Q_\nu^\mu$ ($\mu,\nu = 1,\ldots ,M$)  
are specific functions of $x$, $U$ and $\partial_x U$; 
$v = (v^1,\ldots ,v^M)$ is an arbitrary solution of some linear system of PDEs 
\begin{equation}
\label{eq3_11} 
\tilde {L}^\alpha_\mu[X] v^\mu 
=
\tilde{b}_{\alpha\mu}(X)v^\mu 
+ \tilde{b}_{\alpha\mu i}(X)v^\mu_{X_i} + \cdots 
+ \tilde{b}_{\alpha\mu i_1 \cdots i_k}(X)v^\mu_{X_{i_1}\cdots X_{i_k}} 
= 0,\quad \alpha = 1,\ldots ,m,
\end{equation}
in terms of specific independent variables $X = (X_1 (x,U),\ldots ,X_n (x,U))$; and $J[U] = \vert DX / Dx\vert =\det(DX_i(x,U)/Dx_j)$
which is a function of $x$, $U$ and $\partial_x U$. 
\end{theorem}

\begin{proof}
The existence of multipliers (\ref{eq35}) follows from equation (\ref{eq33}) 
with $X$ playing the role of $z$, 
$\tilde {L}_\nu^\alpha [X]$ playing the role of $L_\nu^{*\alpha} [z]$ 
and with $V^\mu = v^\mu $ satisfying system \eqref{eq3_11}.
\end{proof}

\begin{theorem}[Sufficient conditions for the existence of an invertible 
linearization mapping]
Suppose a given nonlinear system ${\bf R}\{x,u\}$ of PDEs \eqref{eq1} admits 
conservation law multipliers of the form \eqref{eq35} 
where the components of $v$ are dependent variables of some linear system \eqref{eq3_11} 
with specific independent variables $X = (X_1 (x,u),\ldots ,X_n (x,u))$. 
Let $\tilde {L}^* [X]$ be the adjoint of the linear operator $\tilde {L}[X]$
in (\ref{eq3_11})
and consider the augmented system of PDEs
consisting of the given nonlinear system \eqref{eq1} and the linear system \eqref{eq3_11}. 
Then, from identity \eqref{eq33},
there exists an infinite set of multipliers 
\[
\{\Lambda_\mu[U,V], \tilde\Lambda_\alpha[U]\} 
= \{
Q_\mu^\nu(x,U,\partial_x U) \delta_{\lambda\nu}V^\lambda(X(x,U)) J[U],
-\delta_{\alpha\beta} W^\beta(x,U)J[U] 
\}
\]
$(\mu=1,\ldots,M;\alpha=1,\ldots,m)$ 
yielding a conservation law 
\begin{equation}
\label{eq36} 
\Lambda_\nu[U,V] G^\nu [U] 
- \tilde\Lambda_\alpha[U] \tilde{L}_\mu ^\alpha [X(x,U)]V^\mu 
= D\Gamma_i / Dx_i
\end{equation}
for some specific functions $\Gamma_i$, 
in terms of the Jacobian
\begin{equation}
\label{eq37} 
J[U] = \vert DX / Dx\vert  = \det (DX_i(x,U)/Dx_j ).
\end{equation}
This conservation law \eqref{eq36} is equivalent to the identity
\begin{equation}
\label{eq38} 
\delta_{\lambda\mu}V^\lambda Q_\nu^\mu(x,U,\partial_x U) G^\nu [U] 
- \delta_{\alpha\beta}W^\alpha (x,U)\tilde{L}_\mu^\beta [X(x,U)]V^\mu 
= D\Upsilon_i / DX_i ,
\end{equation}
holding for some functions $\Upsilon_i$. 
If the variables $\{X_i\}$ ($i=1,\ldots,n$) are functionally independent
and if the factors $\{Q_\nu^\mu\}$ ($\mu,\nu=1,\ldots,M$) are non-degenerate,
then the point transformation given by
\begin{equation}
\label{eq39} 
z = X(x,u),\quad 
w = W(x,u)
\end{equation}
maps the nonlinear system of PDEs \eqref{eq1} invertibly into the linear system
${\bf S}\{z,w\}$ given by
\beq\label{eq40} 
\barr
\tilde {L}_\alpha ^{*\nu} [z]w^\alpha 
= \displaystyle \tilde b_{\alpha\nu}(z)w^\alpha 
-\frac{\partial}{\partial z_i}\Big(\tilde b_{\alpha\nu i}(z)w^\alpha\Big) + \cdots  
+ ( - 1)^k\frac{\partial ^k}{\partial z_{i_1 } \cdots \partial z_{i_k } }
\Big(\tilde b_{\alpha\nu i_1 \cdots i_k}(z)w^\alpha \Big)
\\\qquad\qquad
= 0,\quad \nu=1,\ldots,M .
\earr\eeq
\end{theorem}

\begin{proof}Since $\tilde {L}[X]$ is a linear operator, the identity (\ref{eq31}) gives 
\begin{equation}
\label{eq41} 
\delta_{\alpha\beta}W^\alpha (x,U)\tilde {L}_\mu^\beta [X(x,U)]V^\mu = 
\delta_{\mu\nu}V^\mu \tilde {L}_\alpha^{*\nu}[X(x,U)]W^\alpha (x,U) + D\Gamma_i / DX_i
\end{equation}
for some specific functions $\Gamma_i [U,V,W]$. 
Consequently, the identity (\ref{eq38}) becomes
\begin{equation}
\label{eq42} 
\delta_{\lambda\mu}V^\lambda (Q_\nu^\mu [U]G^\nu [U] 
- \tilde {L}_\alpha^{*\mu} [X(x,U)]W^\alpha (x,U)) 
= D(\Gamma _i + \Upsilon_i ) / DX_i .
\end{equation}
Now apply the Euler operators with respect to $V^\mu,$ i.e.,
\[
E_{V^\mu } = \frac{\partial }{\partial V^\mu } - \frac{D}{DX_i }\frac{\partial }{\partial V^\mu_{X_i}} + \cdots ,
\quad \mu = 1,\ldots,M,
\]
to each side of equation (\ref{eq42}). 
Each of these Euler operators annihilates the right-hand side of (\ref{eq42}) and hence we obtain the identity
\begin{equation}
\label{eq43} Q_\nu ^\mu [U]G^\nu [U] = \tilde {L}_\alpha^{*\mu} [X(x,U)]W^\alpha (x,U)
\end{equation}
holding for arbitrary functions $U$. 
Now suppose $U = u(x)$ solves the given nonlinear system of PDEs (\ref{eq1}),
then it follows that $w = W(x,u(x))$ solves the linear system given by (\ref{eq40}). 
Consequently, one obtains the point transformation (\ref{eq39})
which maps the nonlinear system of PDEs (\ref{eq1}) 
into the linear system (\ref{eq40}).
Invertibility of this transformation is established by Proposition~3.1. 
\end{proof}

\subsection{Computational Steps}
We now outline the computational steps involved in applying Theorems 3.1 and 3.2 to linearize a given nonlinear system of
PDEs through admitted conservation law multipliers.

\textbf{Step 1:} 
For a given nonlinear system ${\bf R}\{x,u\}$ of PDEs (\ref{eq1}) 
with highest derivative of order $k$, 
solve the determining equations (\ref{deteq}) 
for admitted sets of multipliers $\{\Lambda_\nu[U]\}$ 
depending on the independent variables $x = (x_1 ,\ldots, x_n )$, 
the dependent variables $U = (U^1,\ldots,U^m)$ 
and their first derivatives $\partial_x U = (U^1_{x_i},\ldots,U^m_{x_i})$ . 
Two cases arise, depending on whether ${\bf R}\{x,u\}$ admits any
set of multipliers of the form (\ref{eq35}) with functions $v=(v^1(X), \ldots, v^M(X))$
satisfying some linear system of PDEs of the form \eqref{eq3_11} 
in terms of specific independent variables $X = (X_1(x,U),\ldots ,X_n (x,U))$. 

\textit{Case I. No set of multipliers of the required form are admitted.}
Then one concludes from Theorem 3.1 that 
${\bf R}\{x,u\}$ cannot be mapped invertibly by any point transformation 
to a linear system.

\textit{Case II. There is an infinite set of multipliers with the required form.}
Typically in this case, the variables $X_i(x,U)$ are found directly from 
the specific form of the multiplier determining equations, 
either by inspection (see examples 4.1.2 and 4.2.2) 
or, more generally, through integration of characteristic first-order linear PDEs 
contained in the system of multiplier determining equations (see example 4.3.2). 
Then from the admitted multipliers (\ref{eq35}) one reads off (by Theorem 3.1)
\begin{enumerate}
\item[(a)] 
the independent variable part of the linearizing point transformation \eqref{eq39}, 
i.e. $x_i\rightarrow z_i = X_i(x,u)$;
\item[(b)] 
the specific linear target system which is given by the adjoint of the linear system \eqref{eq3_11}.
\end{enumerate}

\textbf{Step 2:} 
Assuming that the necessary conditions of Theorem 3.1 hold (\textit{Case II}),
one proceeds by replacing $v(X)$ by an arbitrary function $V(X)$ 
in the conservation law (\ref{eq26}) arising from 
the multipliers (\ref{eq35}) for the nonlinear system ${\bf R}\{x,u\}$ of PDEs (\ref{eq1}).
This will automatically yield --- 
after integration by parts if necessary 
(with the use of the transformation formula for divergence expressions 
derived in \cite{f}) --- 
a conservation law identity of the form (\ref{eq38}) 
for the augmented system consisting of the given nonlinear system (\ref{eq1})  
and the linear system (\ref{eq3_11}). 
 From the components of the multipliers for the linear system (\ref{eq3_11})
in this identity, one directly obtains
\begin{enumerate}
\item[(c)] 
the dependent variable part 
$u^\alpha \rightarrow w^\alpha=W^\alpha(x,u)$ 
of the linearization mapping to the linear target system. 
\end{enumerate}

In summary, 
when there exists a linearization of a given nonlinear PDE system 
${\bf R}\{x,u\}$ $(m\geq 1)$
by an invertible point transformation then 
the necessary conditions stated in Theorem~3.1 
yield a linear target system along with the independent variables of this system, 
while the sufficient conditions stated in Theorem~3.2 
yield the dependent variables of the linear target system, 
which completes the transformation. 
(Examples are illustrated in sections~4.1.2 and 4.3.2.)
Furthermore, if the necessary conditions do not hold then 
no linearizing point transformation exists. 

It is easy to extend these results to include 
the linearization of scalar PDEs ($m=1$) by contact transformations,
where $X_i$ and $W^\alpha$ can now depend on first derivatives of $U$. 
Since in this case it follows that 
the Jacobian $J[U]$ and the factors $Q_\mu^\nu[U]$
may depend on second derivatives of $U$, 
one must consequently seek sets of multipliers $\{\Lambda_\nu[U]\}$ 
depending on $x$, $U$, $\partial_x U$ and $\partial_x^2 U$. 
Then in Step 1 any admitted multipliers of the form (\ref{eq35})
will yield the independent variable part of 
a linearizing contact transformation, 
i.e., $x_i \rightarrow z_i = X_i(x,u,\partial_x u)$; 
similarly, in Step 2 the resulting conservation law identity (\ref{eq38})
will yield the dependent variable part of this mapping, 
i.e., $u^\alpha \rightarrow w^\alpha=W^\alpha(x,u,\partial_x u)$. 
(An example is illustrated in section~4.2.2.)
If no such multipliers are admitted then 
no linearizing contact transformation exists.  

We remark that in certain special cases 
a lower differential order is possible in the required form of
conservation law multipliers (\ref{eq35}) 
for the existence of a linearizing transformation. 
Such situations are important computationally
because the multiplier determining equations will then become 
more over-determined and hence easier to solve. 

(1) 
For point or contact transformations that involve 
{\it no change in the independent variables}, 
$x_i \rightarrow z_i = x_i= X_i$, 
one can easily show from the results in \cite{f} that 
the Jacobian (\ref{eq34}) will be trivial,
i.e. $J=\det(DX_i/Dx_j)=\det(\delta_i^j)=1$,
while the factors $Q_\nu^\mu[U]$ in (\ref{eq32}) will have
no dependence on $\partial_x U$ in the multicomponent case ($m\geq 2$)
and no dependence on $\partial_x^2 U$ in the scalar case ($m=1$). 
Hence, in this situation, 
conservation law multipliers can be sought in the form
\[
\Lambda_\nu [U] = \delta_{\lambda\mu}v^\lambda(x) Q_\nu^\mu [U]
\]
depending at most on $x$, $U$ and $\partial_x U$ if $m=1$ 
or only on $x$ and $U$ if $m\geq 2$. 
(A fully algorithmic version of the linearization steps 
applied to this situation was given in \cite{w}, 
which allows one to detect and explicitly find all linearizing mappings 
that involve only a transformation of dependent variables.)
See example 4.1.2. 

(2) 
Suppose a given nonlinear PDE system \eqref{eq1} of order $k$ admits 
{\it integrating factors} 
$\{\lambda_\nu^\mu(x,U)\}$ for its highest derivative terms, 
so that the system has the form 
$\lambda_\nu^\mu[U] G^\nu[U] = D\tilde G^\nu_i[U]/Dx_i + \tilde G^\nu_0[U]$
where $\tilde G_0^\nu$ and $\tilde G^\nu_i$ ($i=1,\ldots,n$)
are of order at most $k-1$ in derivatives of $U$. 
Then it is straightforward to show that the factors relating
the nonlinear and linear PDE systems in (\ref{eq32}) 
will be given by $Q_\nu^\mu [U] = \lambda_\nu^\mu(x,U) J[U]^{-1}$,
and consequently one can seek conservation law multipliers of the form
\[
\Lambda_\nu [U] = \delta_{\lambda\mu}v^\lambda(X) \lambda_\nu^\mu(x,U)
\]
where $X_i[U]$ depends at most on $x$, $U$ and $\partial_x U$ if $m=1$
or only on $x$ and $U$ if $m\geq 2$
(corresponding to linearizing contact transformations or point transformations,
respectively). 
See examples 4.2.2 and 4.3.2.

\section{Examples}\zeroall

To illustrate the conservation law approach 
for obtaining linearizations and contrast it with the symmetry approach, we now consider three examples.

\subsection{Linearization of Burgers' Equation}
Our first example is a nonlinear PDE system 
with independent variables $(x_1 ,x_2 ) = (x,t)$ 
and dependent variables $(u^1,u^2)$ as given by 
\beq\label{eq44}
G^1[u] = \displaystyle \frac{\partial u^2}{\partial x} - 2u^1 = 0,\qquad
G^2[u] = \displaystyle \frac{\partial u^2}{\partial t} - 2\frac{\partial u^1}{\partial x} + (u^1)^2 = 0,
\eeq
where $u^1$ satisfies Burgers' equation
\begin{equation}
\label{eq45} u^1_{xx} - u^1u^1_x - u^1_t = 0.
\end{equation}

\subsubsection{Linearization Through Admitted Point Symmetries}

By a direct computation (solving the symmetry determining equations), 
one finds that equation (\ref{eq45}) admits at most a finite number of contact symmetries. 
Hence there does not exist any contact (or point) transformation 
that linearizes Burgers' equation.

On the other hand the nonlinear system (\ref{eq44}) is found to admit
an infinite set of point symmetries 
given by the infinitesimal generator \cite{b12,b13} 
\begin{equation}
\label{eq46} 
{\bf X} = e^{u^2 / 4}\left( (2g_x(x,t) + g(x,t)u^1)\frac{\partial }{\partial u^1} 
+ 4g(x,t)\frac{\partial }{\partial u^2} \right)
\end{equation}
where $g(x,t)$ is an arbitrary solution of the linear PDE
\beq\label{eq47} 
g_{xx} = g_t .
\eeq
Consequently, one can apply Theorems 2.1 and 2.2 to linearize the system (\ref{eq44}) as follows. 
First we see from a comparison of (\ref{eq46}) with expressions (\ref{eq10}) that 
one has 
$F^1 = g_x(x,t)$, $F^2 = g(x,t)$, 
with 
$\alpha_{ij} = 0$, 
$\beta _1^1 = 2e^{u^2 /4}$, $\beta _2^1 = u^1e^{u^2 / 4}$, 
$\beta _1^2 = 0$, $\beta _2^2 = 4e^{u^2 / 4}$; 
the associated linear homogeneous system (\ref{eq12}) 
has $X_1 = x$, $X_2 = t$ 
as functionally independent solutions 
and the corresponding linear inhomogeneous system (\ref{eq13}) 
has a particular solution 
$(\psi ^1,\psi ^2) = (\frac{1}{2}u^1e^{ - u^2 / 4}, - e^{ - u^2 /4})$. 
Then from (\ref{eq47}) we have that $F=(F^1,F^2)$ satisfies the linear system 
\beq
F^2_x = F^1 ,\quad F^1_x=F^2_t .
\eeq
Hence one obtains the invertible point transformation
\begin{equation}
\label{eq48} 
z_1 = x,\quad 
z_2 = t, \quad
w^1 = \textstyle\frac{1}{2} u^1e^{ - u^2 / 4},\quad 
w^2 = - e^{ - u^2 / 4}
\end{equation}
mapping the given nonlinear system (\ref{eq44}) into the linear system
\beq\label{eq49} 
\frac{\partial w^2}{\partial x}= w^1 ,\quad 
\frac{\partial w^1}{\partial x} = \frac{\partial w^2}{\partial t},
\eeq
where both $w^1$ and $w^2$ consequently satisfy the heat equation
\begin{equation}
\label{eq50} 
\frac{\partial ^2w^1}{\partial x^2} - \frac{\partial w^1}{\partial t} = 0 ,\quad
\frac{\partial ^2w^2}{\partial x^2} - \frac{\partial w^2}{\partial t} = 0 .
\end{equation}

We thus note that the mapping (\ref{eq48}) and the linear system (\ref{eq49})
together yield 
\begin{equation}
\label{eq51} 
u^1 = - \frac{2}{w^2}\frac{\partial w^2}{\partial x}
\end{equation}
which is the well-known Hopf-Cole transformation
that maps any solution of the heat equation (\ref{eq50})
into a solution of Burgers' equation (\ref{eq45}).

\subsubsection{Linearization Through Admitted Conservation Law Multipliers}
A straightforward computation of conservation law multipliers 
$\{\Lambda_1(x,t,U), \Lambda_2(x,t,U)\}$
shows that the nonlinear system (\ref{eq44}) admits 
an infinite set of multipliers given by
\begin{equation}
\label{eq52} 
\Lambda_1 = \textstyle\frac{1}{2}U^1e^{- U^2/4}f(x,t) + e^{- U^2/4}f_x(x,t) ,\quad 
\Lambda_2 = e^{ - U^2/4}f(x,t), 
\end{equation}
where $f(x,t)$ is an arbitrary solution of the linear scalar PDE
$f_{xx}+f_t=0$
(namely, the backward heat equation). 
Hence, the necessary conditions of Theorem~3.1 
for the existence of an invertible mapping of the nonlinear system (\ref{eq44}) 
to a linear target system are satisfied, where, 
through a comparison of (\ref{eq52}) with (\ref{eq35}), 
the adjoint target system is given by 
\beq\label{eq53} 
\frac{\partial v^1}{\partial x} - v^2 = 0, \quad
\frac{\partial v^2}{\partial x} + \frac{\partial v^1}{\partial t} = 0
\eeq
with dependent variables 
\beq\label{eq53b}
v=(v^1,v^2)=(f,f_x)
\eeq
and the same independent variables $(x,t)$ as the given system (\ref{eq44}). 

In the conservation law (\ref{eq26}) arising from 
the multipliers (\ref{eq52}), (\ref{eq53}) and (\ref{eq53b}),  
we replace $(v^1(x,t),v^2(x,t))$ by arbitrary functions $(V^1(x,t),V^2(x,t))$. 
This leads to the following conservation law identity for the augmented system consisting of the given nonlinear system
(\ref{eq44}) and the linear system (\ref{eq53}):
\beq\label{eq54} \barr
\displaystyle 
\left(V^1(\textstyle\frac{1}{2}U^1e^{ - U^2 / 4}) + V^2e^{ - U^2 / 4}\right)G^1[U] + V^1e^{ - U^2 / 4}G^2[U]
\\[2ex]
\displaystyle  
- 2U^1e^{ - U^2 / 4}\left(\frac{\partial V^1}{\partial x} - V^2 \right) 
- 4e^{ - U^2 / 4}\left(\frac{\partial V^2}{\partial x} + \frac{\partial V^1}{\partial t} \right)
\\[2ex]
 = \displaystyle 
\frac{D}{Dx}\left(e^{ - U^2 / 4}( - 4V^2 - 2U^1V^1) \right) +
\frac{D}{Dt}\left(- 4V^1e^{ - U^2 / 4} \right).
\earr\eeq
Consequently, 
directly comparing (\ref{eq54}) with the identity (\ref{eq38}) in Theorem 3.2, 
we obtain an invertible mapping 
$(u^1,u^2) \rightarrow (w^1=2u^1e^{-u^2/4}, w^2=4e^{-u^2/4})$
of the nonlinear system (\ref{eq44}) 
to a linear target system given by the adjoint of the linear system (\ref{eq53}). 
In particular, this yields the point transformation (\ref{eq48}) 
that maps the nonlinear system (\ref{eq44}) into the linear system (\ref{eq49})
up to a suitable rescaling of the dependent variables $w=(w^1,w^2)$.

\subsection{Linearization of a pipeline flow equation}
As a second example, we consider the pipeline flow equation \cite{b11}
\begin{equation}
\label{eq55} 
G[u] = u_t u_{xx} + u_x{}^p = 0, \quad p={\rm const}.
\end{equation}

\subsubsection{Linearization Through Admitted Contact Symmetries}
The nonlinear scalar PDE (\ref{eq55}) admits an infinite set of contact symmetries 
whose infinitesimal generator is given by 
\begin{equation}
\label{eq56} 
{\bf X} = - \frac{\partial F}{\partial u_x }\frac{\partial }{\partial x} + \left( {F - u_x \frac{\partial F}{\partial
u_x }} \right)\frac{\partial }{\partial u} + \frac{\partial F}{\partial t}\frac{\partial }{\partial u_t }
\end{equation}
where $F = F(t,u_x )$ is an arbitrary solution of the second-order linear PDE
\begin{equation}
\label{eq57} 
u_x{}^p \frac{\partial ^2F}{\partial u_x{}^2 } - \frac{\partial F}{\partial t} = 0.
\end{equation}
Consequently, one can linearize the PDE (\ref{eq55})
by applying the generalization of Theorems~2.1 and~2.2
for seeking a linearizing contact transformation. 
In particular, 
$X_1 = u_x$, $X_2 = t$ are found to be functionally independent solutions
of the appropriate generalization (cf. \cite{b01,b02}) 
of the linear homogeneous system (\ref{eq12}),
while the corresponding generalization of the linear inhomogeneous system (\ref{eq13}) 
has the particular solution
$\psi = u - xu_x$ 
(from which one deduces $(\rho^1,\rho^2) = (u_t,-x)$ through 
the contact condition in Theorem~1.2).
Consequently, one obtains an invertible mapping $\mu$ 
given by the contact transformation
\begin{equation}
\label{eq58} z_1 = t,\quad z_2 = u_x ,\quad w = u - xu_x ,\quad w_{z_1} = u_t ,\quad w_{z_2} = - x,
\end{equation}
which maps the nonlinear PDE (\ref{eq55}) to the linear PDE given by
\begin{equation}
\label{eq59} z_2^{\ p} \frac{\partial ^2w}{\partial z_2^{\ 2}} - \frac{\partial w}{\partial z_1} = 0.
\end{equation}

\subsubsection{Linearization Through Admitted Conservation Law Multipliers}

Through a computation of conservation law multipliers $\Lambda(x,t,U,U_x,U_t)$,
one finds that the given nonlinear PDE (\ref{eq55}) admits 
an infinite set of multipliers 
\begin{equation}
\label{eq60} 
\Lambda = v(U_x,t),
\end{equation}
where this function satisfies the linear scalar PDE 
\[ 
v_t + (U_x{}^p v)_{U_x U_x} = 0. 
\]
By inspection, since $v$ depends on two variables 
(i.e., the same number of independent variables as $U$ has), 
we see that the necessary conditions of Theorem~3.1 
for the existence of an invertible contact transformation that linearizes 
the nonlinear scalar PDE (\ref{eq55}) are satisfied 
with the adjoint target system given by
\beq
\label{eq61} 
\frac{\partial v}{\partial T} + \frac{\partial ^2(X^pv)}{\partial X^2} = 0
\eeq
for $v=v(X,T)$ in terms of the independent variables
\beq
\label{eq61b}
X = U_x,\quad T = t .
\eeq

In the conservation law (\ref{eq26}) arising from the multipliers (\ref{eq60}), 
we replace $v(X,T)$ by an arbitrary function $V(X,T)$. 
This leads to the following conservation law identity for the augmented system 
consisting of the given nonlinear PDE (\ref{eq55}) and the linear PDE (\ref{eq61}):
\beq\label{eq62} \barr
\displaystyle 
VG[U] - (xU_x - U)J[U]\left(\frac{\partial V}{\partial T} + 
\frac{\partial ^2(X^pV)}{\partial X^2}\right)
\\[2ex]
 = \displaystyle
\frac{D}{Dx}\Big((xU_x - U)(U_{tx} V + U_x ^pV_X ) + \big((1 - p)xU_x
+ pU\big)U_x ^{p - 1}V \Big)
+ \frac{D}{Dt}\Big(U_{xx} (U - xU_x )V \Big),
\earr\eeq
where 
\begin{equation*}
J[U] = \left| {\frac{D(X,T)}{D(x,t)}} \right| 
= U_{xx} 
\end{equation*}
is the Jacobian of (\ref{eq61b}). 
(For verifying the identity (\ref{eq62}), we note that 
$V_x = V_X U_{xx}$, $V_t = V_X U_{xt} + V_T $.)
Consequently, the sufficiency conditions of Theorem 3.2 hold 
for the existence of an invertible mapping by a contact transformation 
of the nonlinear PDE (\ref{eq55}) to a linear target PDE 
which is the adjoint of the linear PDE (\ref{eq61}). 
In particular, 
from a comparison of (\ref{eq62}) with the identity (\ref{eq38}), 
it follows that the contact transformation 
determined by
\begin{equation}
\label{eq64} 
z_1 = X = u_x ,\quad
z_2 = T = t,\quad
w = xu_x - u 
\end{equation}
maps the nonlinear pipeline equation (\ref{eq55}) invertibly into the linear PDE
\begin{equation}
\label{eq65} 
X^p\frac{\partial ^2w}{\partial X^2} - \frac{\partial w}{\partial T} = 0.
\end{equation}

\subsection{Linearization of a Nonlinear Telegraph Equation}

For a final example, we consider a nonlinear telegraph system given by
\beq\label{tele1}
G^1[u] = \displaystyle
\frac{\partial u^2}{\partial t}-\frac{\partial u^1}{\partial x} = 0,\qquad
G^2[u] = \displaystyle
\frac{\partial u^1}{\partial t}+u^1(u^1-1)-(u^1)^2\frac{\partial u^2}{\partial x} = 0 
\eeq
with dependent variables $(u^1,u^2)$ and independent variables $(x_1,x_2)=(x,t)$.
Note that $u^1$ satisfies the nonlinear telegraph equation
\[ \frac{\partial^2 u^1}{\partial x^{\,2}}-
\frac{\partial}{\partial t}\left((u^1)^{-2}\frac{\partial u^1}{\partial t}+1-(u^1)^{-1}\right) = 0.\]

\subsubsection{Linearization Through Admitted Point Symmetries}

The nonlinear system (\ref{tele1}) admits an infinite set of point symmetries 
given by the infinitesimal generator
\begin{equation}
{\bf X}=
F^1(X,T)\frac{\partial}{\partial x}+e^{-t}F^2(X,T)\frac{\partial}{\partial t}
+e^{-t}u^1 F^2(X,T)\frac{\partial}{\partial u^1}+F^1(X,T)\frac{\partial}{\partial u^2} \label{tele2}
\end{equation}
where 
\begin{equation}
X_1=X=x-u^2,\quad X_2=T=t-\log u^1, \label{tele3}
\end{equation}
and $(F^1(X,T), F^2(X,T))$ is an arbitrary solution of the linear system of PDEs
\begin{equation}
\frac{\partial F^2}{\partial T} - e^T\frac{\partial F^1}{\partial X} = 0 ,\quad
\frac{\partial F^2}{\partial X} - e^T\frac{\partial F^1}{\partial T} = 0 . 
\label{tele4}
\end{equation}
Consequently, one can apply Theorems 2.1 and 2.2 to linearize (\ref{tele1}).  
In particular, 
we see that by comparing (\ref{tele2}) with expressions (\ref{eq10}), 
one has 
$\alpha_{11}=\beta_1^2=1$, $\alpha_{12}=\alpha_{21}=\beta_1^1=\beta_2^2=0$, 
$\alpha_{22}=e^{-t}, \beta_2^1=e^{-t}u$;
the associated linear homogeneous system (\ref{eq11}) 
has $X_1=x-u^2, X_2=t-\log u^1$ as functionally independent solutions 
and the corresponding linear inhomogeneous system (\ref{eq12}) 
has a particular solution  $(\psi^1,\psi^2)=(x,e^t)$.
Hence the mapping given by
\begin{equation}
\label{tele5}
z_1=x-u^2 ,\quad 
z_2=t-\log u^1 ,\quad 
w^1=x ,\quad 
w^2=e^t    
\end{equation}
is an invertible point transformation under which 
the nonlinear telegraph system (\ref{tele1}) is mapped to the linear PDE system
given by
\begin{equation}
\frac{\partial w^2}{\partial z_2} - e^{z_2}\frac{\partial w^1}{\partial z_1} 
= 0 ,\quad
\frac{\partial w^2}{\partial z_1} - e^{z_2}\frac{\partial w^1}{\partial z_2} 
= 0. 
\label{tele6}
\end{equation}	 

\subsubsection{Linearization Through Admitted Conservation Law Multipliers}

Through a computation of conservation law multipliers of the form 
$\{\Lambda_1(x,t,U^1,U^2)$, $\Lambda_2(x,t,U^1,U^2)\}$, 
one can show that the nonlinear telegraph system of PDEs (\ref{tele1}) 
admits an infinite set of multipliers yielding a linearization as follows. 
After some integrability analysis of the multiplier determining equations 
(e.g. using {\sc Crack} \cite{b14}), one first obtains
\begin{equation}
\Lambda_1=f_{U^2},\quad \Lambda_2=f_{U^1}, 
\label{tele7}
\end{equation}
in terms of a function $f(x,t,U^1,U^2)$ satisfying the three linear PDEs
\begin{eqnarray}
&& 
f_x+f_{U^2}=0,\quad 
f_t+U^1f_{U^1}=0,
\label{tele8a} \\
&&
(U^1)^2f_{U^1U^1}+2U^1f_{U^1}-f_{U^2U^2}=0.
\label{tele8b} 
\end{eqnarray}
To proceed we integrate the pair of first-order PDEs (\ref{tele8a}),
which yields a reduction of the independent variables such that 
$f=f(X,T)$ where  
\begin{equation}
\label{tele8c}
X=x-U^2,\quad T=t-\log U^1. 
\end{equation}
Then the second-order PDE (\ref{tele8b}) combined with (\ref{tele7}) 
gives the infinite set of multipliers 
\beq
\Lambda_1=-f_X(X,T),\quad \Lambda_2=-f_T(X,T)/U^1,\quad
f_{XX} -f_{TT} +f_T =0 . 
\label{tele9}
\eeq
Finally, comparing (\ref{tele9}) with (\ref{eq35}) in Theorem~3.1, 
we see that the necessary conditions for the existence of an invertible mapping of
the nonlinear system (\ref{tele1}) to a linear target system are satisfied, 
with the adjoint target system being given by 
\beq
v=(v^1,v^2) =(-f_X,-f_T) . 
\label{tele10a}
\eeq
and
\beq
\frac{\partial v^1}{\partial X} - \frac{\partial v^2}{\partial T} + v^2 = 0 , 
\quad
\frac{\partial v^2}{\partial X} - \frac{\partial v^1}{\partial T}   = 0. 
\label{tele10}
\eeq

In the conservation law (\ref{eq26}) arising from the multipliers (\ref{tele9}), 
(\ref{tele10a}) and (\ref{tele10}), 
we replace $(v^1(X,T),v^2(X,T))$ by arbitrary functions $(V^1(X,T),V^2(X,T))$.
This leads to the following conservation law identity for the augmented system 
consisting of the given nonlinear system (\ref{tele1}) 
and the linear system (\ref{tele10}):
\beq\label{tele11}
\barr
V^1G^1[U]+V^2(U^1)^{-1}G^2[U] 
\displaystyle
- U^1J[U]\left(\frac{\partial V^1}{\partial X}
-\frac{\partial V^2}{\partial T}+V^2\right) 
-xJ[U]\left(\frac{\partial V^2}{\partial X}-\frac{\partial V^1}{\partial T}\right)\\\displaystyle
= 
\frac{D}{Dx}\left(-V^1\left(x\frac{\partial U^2}{\partial t}+\frac{\partial U^1}{\partial t}-U^1\right)
+V^2\left(x-x(U^1)^{-1}\frac{\partial U^1}{\partial t}
-U^1\frac{\partial U^2}{\partial t}\right)
\right)
\\\qquad\displaystyle
+ \frac{D}{Dt}\left(-V^1\left(x-x\frac{\partial U^2}{\partial x}+\frac{\partial U^1}{\partial t}\right)
+V^2\left(x(U^1)^{-1}\frac{\partial U^1}{\partial x}+U^1\frac{\partial U^2}{\partial x}-U^1\right)\right),  
\earr\eeq
where, from (\ref{tele8c}), 
\[ 
J[U]= \left| {\frac{D(X,T)}{D(x,t)}} \right| 
= (U^1)^{-1}\left( \left(1-\frac{\partial U^2}{\partial x}\right)\left(U^1-\frac{\partial U^1}{\partial t}\right) 
-\frac{\partial U^2}{\partial t}\frac{\partial U^1}{\partial x}\right). 
\]
Consequently, 
by comparing (\ref{tele11}) with the identity (\ref{eq38}) in Theorem 3.2,
we obtain an invertible mapping of the nonlinear system (\ref{tele1}) 
to a linear target system 
which is given by the adjoint of the linear system (\ref{tele10}). 
In particular, the point transformation
\begin{equation}
z_1 = X = x-u^2,\quad 
z_2 = T = t-\log u^1,\quad 
w^1 = x,\quad 
w^2 = u^1   
\label{tele12}
\end{equation}
maps the nonlinear telegraph system (\ref{tele1}) invertibly into the linear system 
\begin{equation}
\frac{\partial w^1}{\partial X} - \frac{\partial w^2}{\partial T} - w^2 = 0 ,\quad
\frac{\partial w^2}{\partial X} - \frac{\partial w^1}{\partial T} = 0 .
\label{tele13}
\end{equation}    
Note that the further point transformation
$\tilde{w}^1 = w^1$, $\tilde{w}^2 = e^Tw^2$
maps this linear system (\ref{tele13}) into 
the equivalent linear system (\ref{tele6}) obtained from
linearization of the nonlinear telegraph system (\ref{tele1}) through
its admitted infinite set of point symmetries (\ref{tele2}).

\section{Concluding Remarks}

An important question is which of the two methods presented in this paper
is better for linearization? 
It would appear likely that the approach using 
admitted conservation law multipliers 
will be better computationally 
since, in general, 
the solution space for the determining equations for conservation law multipliers 
is smaller than that for point (or contact) symmetries. 
A way of seeing this is to consider 
the case when a given system of PDEs is variational, 
i.e., its linearization operator (Frechet derivative) is self-adjoint,
so then Noether's theorem is applicable. 
For any such system 
its conservation law multipliers correspond to a variational subset of 
its admitted point (or contact) symmetries 
and hence, in general, 
the determining equations for conservation law multipliers are then 
more over-determined than those for point (or contact) symmetries 
since they include the symmetry determining equations as a subset 
\cite{b05,b06,b07}. 

An interesting computational way for checking for the possibility of linearization of a given system of nonlinear PDEs, 
especially for classification problems, 
is to apply the work in \cite{c,d} 
to find the size of the solution space of a given system of 
determining equations
without the need for obtaining any of its actual solutions.  
In particular, if the size of the solution space of the determining systems 
either for conservation law multipliers 
or for point (or contact) symmetries is finite dimensional, 
then no linearization by an invertible mapping is possible. 
But the converse (that the solution space is infinite dimensional) 
is not sufficient to imply the existence of a linearization by
a point or contact transformation. 
Specifically, 
the solution space must have a sufficiently large number of parameters
such that the number of functions and independent variables, 
as well as the number of linear PDEs, 
matches the corresponding number in the given nonlinear system of PDEs. 
This counting can be performed fully algorithmically \cite{c,d}
provided a differential Gr\"{o}bner basis is available. 
However, it is important to recognize that in practice 
the computational complexity of finding a differential Gr\"{o}bner basis 
will be very high. 

It would also be interesting to develop a fully algorithmic version 
of the two linearization methods themselves. 
This would entail, firstly, 
the use of a differential Gr\"{o}bner basis computation 
to reduce the determining equations into a consistent over-determined linear system 
satisfying all integrability conditions 
and having a sufficiently large solution space. 
Secondly, a strictly computational procedure must be specified to extract
from this linear system the form of the linearizing transformation for 
the dependent and independent variables in the given nonlinear system of PDEs. 

The algorithmic steps we have used in our present work 
(e.g. for linearizing Burgers' equation, the pipeline equation, 
and the nonlinear telegraph system) 
are less automatic than finding a differential Gr\"{o}bner basis 
but have much less computational complexity in practice. 
In particular, they take advantage of the integration capabilities of {\sc Crack} \cite{b14}, 
including algorithms for direct and indirect separation of equations, 
for dropping redundant functions, for checking integrability conditions, 
and others. 
Furthermore, {\sc Crack} also handles solving the first-order linear PDEs 
as needed in finding the transformations of the independent variables 
in the linearization mapping. 

Finally, we mention that a hodograph-type approach for linearization
has been given some time ago in work \cite{e} of Varley and Seymour 
on linearizing the nonlinear telegraph system (example 4.3). 
Their clever procedure for finding a linearization is not symmetry-based; 
it only applies to specific types of nonlinear PDEs 
and cannot determine, in general, 
whether a given system of nonlinear PDEs admits a linearization 
by an invertible mapping.

\section*{Acknowledgements}
The authors are each supported by an NSERC research grant.

\end{document}